\documentclass[a4paper,12pt]{article}
\usepackage[dvips]{graphicx}

\usepackage{amsmath}

\begin{document}


\title{
Phase Transitions 
in one-dimensional nonequilibrium  systems
}
\author{ M.\ R.\ Evans \\[2ex]
	Department of Physics and Astronomy,\\
	The University of Edinburgh,\\
	Mayfield Road,\\
	Edinburgh EH9 3JZ,\\
	U.K.}
\date{21st December, 1999}
\maketitle

\begin{abstract}

The phenomenon of phase transitions in one-dimensional systems is
discussed. Equilibrium systems are reviewed and some properties of an
energy function which may allow phase transitions and phase ordering
in one dimension are identified.  We then give an overview of the
one-dimensional phase transitions which have been studied in
nonequilibrium systems.  A particularly simple model, the zero-range
process, for which the steady state is known exactly as a product
measure, is discussed in some detail.  Generalisations of the model,
for which a product measure still holds, are also discussed.  We
analyse in detail a condensation phase transition in the model and
show how conditions under which it may occur may be related to the
existence of an effective long-range energy function.  It is also
shown that even when the conditions for condensation are not fulfilled
one can still observe very sharp crossover behaviour and apparent
condensation on a finite system.  Although the zero-range process is
not well known within the physics community, several nonequilibrium
models have been proposed that are examples of a zero-range process,
or closely related to it, and we review these applications here.

\end{abstract}


\section{Introduction}

In recent years the study of nonequilibrium systems has come to the
fore in statistical mechanics. Basically, one can consider two types
of nonequilibrium systems: those relaxing towards thermal equilibrium
and those held far from thermal equilibrium {\it e.g.} by the system
being driven by some external field. In the present article we will be
mainly concerned with the latter scenario.

To be more specific we define our nonequilibrium systems as those
evolving through a local stochastic dynamics which {\it a priori} does
not obey detailed balance, at least not with respect to any
`reasonable' energy function.  The question of what is a reasonable
energy function is a moot point.  One might propose that the energy
contains only local interactions, or is extensive, or is written down
according to some physical principles; but any answer to the question
is subjective.  However, the basic point is that the nonequilibrium
system is {\em defined} by its dynamics without regard to any concept
of energy and it is the dynamics which should seem reasonable or
`physical'. This is distinct from an equilibrium system where the
energy function should be `physical' and the dynamics is usually
defined in an ad hoc way simply to guarantee that one obtains the
Gibbs-Boltzmann weight with the specified energy. The easiest way to
do this is to use the detailed balance condition.

A natural way to construct a nonequilibrium steady state is to drive
the system by forcing a current of some conserved quantity, for
example energy or mass, through the system.  Such systems are known as
driven diffusive systems (DDS).  The archetypal model was introduced
Katz, Lebowitz and Spohn \cite{KLS}. Basically it comprises a two
dimensional Ising-like lattice gas evolving under conservative
Kawasaki dynamics (spin exchange) and with a drive direction
imposed. It has been shown that a continuous phase transition
exists in the driven system,
as is also the case  in the undriven (Ising) system,
but, most interestingly, one sees {\it generic} long range (power-law
decay) correlations as opposed to the undriven systems where
long-range correlations are only seen at criticality.  Although exact
results are not available for this system\cite{SZ}, it is often
thought that generic power-law correlations are related to the
existence of an effective long-range Hamiltonian for the system (see
{\it e.g.} \cite{BRX, AE}).

More recently it has been realised that DDS in one dimension exhibit
non-trivial behaviour. The interest has been from a fundamental
viewpoint but also in the context of applications such as interface
growth \cite{Krug97} and traffic flow modelling \cite{SW}. Also it
turns out that problems of transport with a single-file constraint
have long been of interest in biological contexts such as transport
across membranes \cite{Heckmann,CL} and the kinetics of
biopolymerisation\cite{MGP}.

One intriguing feature of one-dimensional systems is the possibility
of phase ordering and phase transitions. In recent years this
possibility has begun to be explored and some examples are by now well
studied.  To appreciate the significance one should recall the general
dictum that in one-dimensional equilibrium systems phase ordering and
phase transitions do not occur (except in the limit of
zero-temperature, or with long range interaction---see section 2).  In
the one-dimensional nonequilibrium systems studied so far it appears
that the presence of conserved quantities and an imposed drive are
important in allowing ordering and phase transitions.  However there
still does not exist a general theoretical framework within which to
understand the phenomena.

The purpose of this article is twofold.  Firstly I wish to give a
broad overview of phase transitions and phase ordering in one
dimension---this is carried out in Section~ \ref{Sec:PT1d}.  In
particular, in Section~\ref{Sec:equil} we discuss the conditions under
which ordering and phase transitions may occur in equilibrium systems
{\it i.e.} the requisite properties of the energy function to allow
such phenomena. Then in Section~\ref{Sec:nonequil} we catalogue some
nonequilibrium, one dimensional systems which exhibit non-trivial
phase behaviour

The second purpose is to discuss a very simple class of microscopic
models, the zero-range processes \cite{Spitzer,Andjel}, which are
presented in section~\ref{Sec:ZRP}. For these models the steady state
can be calculated exactly since it factorises into a product
measure. There is some irony in the fact that the system has found
widespread application in the modelling of nonequilibrium phenomena
(see Section~\ref{Sec:app}), although the zero-range process was
originally introduced by Spitzer\cite{Spitzer} as a dynamics which
could lead to Gibbs measures.  In section~4 we discuss generalisations
of the basic model which also have a product measure steady state. We
show in Section~{\ref{Sec:cond} how the model can exhibit a phase
transition, that we shall refer to as a condensation transition, which
is analysed in some detail.  We also discuss an interesting sharp
crossover phenomenon whereby models, although not fulfilling the
conditions for strict condensation and phase ordering, may often
appear to be in a condensed phase on a finite system.  The simplicity
of the system allows us to explore the roles of a conserved quantity,
the presence of a drive and effective long-range energy
functions. Conclusions are drawn in Section~\ref{Sec:Conc}.

\section{Phase transitions in one dimension}
\label{Sec:PT1d}
\subsection{One-dimensional equilibrium systems}
\label{Sec:equil}
As mentioned above, it is received wisdom that in 
one-dimensional equilibrium systems
phase transitions do not occur.  In fact any careful statement of this
requires a few caveats and, indeed, a general rigorous statement is
hard to formulate (see \cite{LM} for a discussion).

Perhaps the best known argument is that of Landau and Lifshitz
\cite{LL}. For simplicity, consider a one dimensional lattice of $L$
sites with two possible states, say $A$ and $B$, for each site
variable.  Let us assume the ordered phases, where all sites take
state $A$ or all sites take state $B$, have the lowest energy, and
assume a domain wall (a bond on the lattice which divides a region of
$A$ phase from that of $B$) costs a finite amount of energy
$\epsilon$.  Then $n$ domain walls will cost energy $n \epsilon$ but
the entropic contribution to the free energy due to the number or ways
of placing $n$ walls on $L$ sites $\simeq nT \left[ \ln (n/L) -1
\right]$ $\mbox{for}\quad 1 \ll n \ll L$. Thus for any finite
temperature a balance between energy and entropy ensures that the
number of domain walls grows until it scales as $L$, that is, until
the typical ordered domain size is finite.

Note that this argument relies on a finite energy cost for domain
walls, and short range interactions so that one may ignore the
interaction energy of domain walls.  Indeed, the Ising model with
long-range interactions decaying with distance as $J(r)\sim
r^{-1{-}\sigma}$ has been well studied \cite{FMN} and it has been
demonstrated that the one-dimensional system orders at low
temperatures for $\sigma \leq 1$ \cite{Dyson}.  Also, of course we
require non-zero temperature so that entropy comes into play,
otherwise the two fully ordered states (ground states) would dominate
the partition sum and the system would be frozen into them.

Another even simpler way of thinking of this is from a dynamical
perspective. For a disordered state to order, domain walls
must annihilate each other. However in one dimension no energy is
gained by the two domain walls at opposite ends of a domain moving
closer to one another; a domain always has two domain walls costing
energy $2\epsilon$ no matter what its size is.
Therefore there is no effective force to eliminate domains
and the system is disordered.  Again, this argument
requires a short range interaction so that one can ignore the energy
of interaction of domain walls above some finite distance.

A more mathematical way of addressing the question of phase transition
in 1d is to use the transfer matrix technique \cite{Goldenfeld}. For example,
 on a periodic one-dimensional homogeneous system of $N$
sites, the
partition sum can be written as the trace of a product of $N$ transfer
matrices $T$ :
\begin{equation}
Z = \mbox{Trace} \left[ T^N \right] = \sum_{\lambda} \lambda^N
\end{equation}
where $\lambda$ are the eigenvalues of the transfer matrix.  Now, since
the transfer matrix is finite and the entries are all positive the
Perron-Frobenius theorem \cite{KT} tells us that the largest eigenvalue
$\lambda_{\rm max}$
is non-degenerate.  Thus, there can be no crossing of the largest
eigenvalue as we vary some control parameter. Consequently the free
energy $F \propto \lim_{N{\to} \infty} (\ln Z) / N = \lambda_{\rm max}$
is analytic and we have no phase transitions (which would be signalled
by some non-analyticity of the free energy).

Again, there are exceptions to this reasoning {\it i.e.}  when the
Perron-Frobenius theorem no longer applies.  This can occur when the
transfer matrix becomes infinite due, for example, either to long
range interactions or when the local degree of freedom at each lattice
site is not restricted to a finite number of states {\it e.g.}
\cite{vLH}.  (An extreme instance of the latter case is when we are
actually considering a two dimensional system!)  Another case when the
Perron-Frobenius theorem does not apply is when the transfer matrix
becomes reducible i.e. when there exists components of $T^N$ that are
zero for all values of $N$. This can occur at zero temperature or when
some interaction strengths are set to infinity, an example being the
first order transition in the KDP model discussed in \cite{Nagle}.

In this section we have discussed three arguments, 
presented here at different (low)
levels of rigour, which all point to phase transitions in equilibrium
one-dimensional systems only being possible in the case of long-range
interactions, zero-temperature limit or infinite interaction energies,
or unbounded local variable at a site. As we shall see the situation
for nonequilibrium systems is less restrictive although some
parallels can be drawn.

\subsection{One-dimensional nonequilibrium systems}
\label{Sec:nonequil}
Here we give an overview of one-dimensional systems where phase
transitions and phase ordering may occur.  We focus our attention on
hopping particle models that, despite their simplicity, offer a wide
range of non-trivial behaviour.

A simple one-dimensional model of a driven diffusive system is the
asymmetric simple exclusion process (ASEP). Here particles hop in a
preferred direction on a one-dimensional lattice with hard-core
exclusion (at most one particle can be at any given site).  Indicating
the presence of a particle by a 1 and an empty site (hole) by 0 the
dynamics comprises the following exchanges at nearest neighbour sites
\begin{eqnarray}
1\ 0 &\to& 0\ 1 \quad\mbox{with rate}\quad 1 \nonumber \\
0\ 1 &\to& 1\ 0 \quad\mbox{with rate}\quad q
\end{eqnarray}

The open system was studied by Krug\cite{Krug91} and boundary
induced phase transitions shown to be possible.  Specifically one
considers a lattice of $N$ sites where at the left boundary site (site
1) a particle is introduced with rate $\alpha$ if that site is
empty, and at the right boundary site (site $N$)
any particle present is removed with rate $\beta$. Thus the dynamical
processes at the boundaries are
\begin{eqnarray}
\mbox{ at site $1$} \quad  0 &\to& 1 \quad\mbox{with rate}\quad \alpha \nonumber \\
\mbox{ at site $N$} \quad  1 &\to& 0 \quad\mbox{with rate}\quad \beta\;.
\end{eqnarray}
These boundary conditions force a steady state current of particles
$J$ through the system. Phase transitions occur when
$\lim_{N{\to}\infty} J$ exhibits non-analyticities.  The steady state
of this system was solved exactly for the totally asymmetric case
\cite{DEHP,SD} and more recently for the general $q$ case
\cite{Sasamoto,BECE}.  When $q<1$ the phase diagram comprises three
phases: a high-density phase where the current is limited by a low
exit rate $\beta$ and takes the expression $J=\beta(1-q-\beta)/(1-q)$;
a low-density phase where the current is limited by a low injection
rate $\alpha$ and takes the expression $J=\alpha(1-q-\alpha)/(1-q)$; a
maximal-current phase where both $\alpha,\beta > (1-q)/2$ and the
current is $J=(1-q)/4$. In the maximal current phase generic
long-range correlations exist, an example being the decay of particle
density from the left boundary to the bulk value $1/2$ which is a
power law $\sim 1/x^{1/2}$ where $x$ is distance from the left
boundary.

Clearly the presence of a conserved quantity and a drive, leading to
non-zero current $J$ is crucial to the phase transition. Indeed, the
qualitative phase diagram appears robust for stochastic
one-dimensional driven systems \cite{KSKS}.  For the case of no bulk
drive $q=1$ \cite{SS,SMW}, or `reverse bias' $q>1$ \cite{BECE} the
current vanishes with increasing system size and there are no
boundary-induced phase transitions.

The model has been generalised to two oppositely moving species of
particle: one species is injected at the left, moves rightwards and
exits at the right; the other species is injected at the right, moves
leftwards and exits at the left \cite{EFGM}. Spontaneous symmetry
breaking has been shown to occur, whereby for low exit rates ($\beta$)
the lattice is dominated by one of the species at any given time. In
the low $\beta$ limit the mean flip time between the two
symmetry-related states has been calculated analytically and shown to
diverge exponentially with system size \cite{GLEMSS}.

In these models the open boundaries can be thought of as inhomogeneities
where the order parameter (particle density) is not conserved.
Inhomogeneities which conserve
 the order parameter can be considered on a periodic system. Indeed
a single defect bond on the lattice (through which particles hop
more slowly) is sufficient to cause the system to separate into
two macroscopic regions of different densities \cite{JL}:
a high density region which can be thought of as a traffic jam
behind the defect   and a  low density region in front of the defect.
Here the presence of the drive appears necessary for the defect
to induce the phase separation.

Moving defects ({\it i.e.} particles with dynamics different from that
of the others) have also been considered and exact solutions obtained
\cite{DJLS,TZ,Mallick,DE99,Sasamoto2}. In the model studied in
\cite{Mallick,DE99,Sasamoto2}, varying the rate at which the defect
particle hops forward, denoted $\alpha$, and the rate at which it is
overtaken and exchanges places with normal particles, denoted $\beta$,
produces a phase diagram closely related to the open boundary problem.
Moreover for low $\beta$ and high $\alpha$ there is a phase where the
defect particle induces phase separation between a high density region
behind it and a low density region in front of it.

For some of the examples discussed so far the steady state has been
solved exactly by constructing a matrix product which is reviewed
in \cite{Derrida98}.  This
reveals that the steady state weights are very complicated functions
of the particle number and positions. It does not appear easy to
relate this to any concept of an energy function.  Indeed, it has been
shown that a matrix product state is non-Gibbsian \cite{Speer}.

A natural question to ask is whether systems related to the hopping
particle models described so far, but without inhomogeneities, can
exhibit phase ordering. A very simple model was introduced in
\cite{EKKM} comprising three species of conserved particles, amongst
which all possible exchanges are allowed. However a key feature is
that the dynamics has a cyclic symmetry.  To be specific let each site
of a one-dimensional periodic lattice be occupied either by an $A, B$
or $C$ particle (there are no holes in this model).  The dynamical
exchanges are
\begin{eqnarray}
A\ B &\to& B\ A \quad\mbox{with rate}\quad q \nonumber \\
B\ A &\to& A\ B \quad\mbox{with rate}\quad 1 \nonumber \\
B\ C &\to& C\ B \quad\mbox{with rate}\quad q \nonumber \\
C\ B &\to& B\ C \quad\mbox{with rate}\quad 1 \nonumber \\
C\ A &\to& A\ C \quad\mbox{with rate}\quad q \nonumber \\
A\ C &\to& C\ A \quad\mbox{with rate}\quad 1 
\label{ABC}
\end{eqnarray}
and we will take $q<1$. For example, the hopping of an $A$ particle
is biased to the right when it is an environment of $C$s
and it is biased to the left when it is in an environment of $B$s.

The phase separation observed in the model is rather easy to understand:
if the system has separated into a domain of $A$s, followed by a domain of
$B$s, followed by a domain of $C$s (in that order), then the domain
walls that are present $AB$, $BC$, $CA$ are all stable objects.
This is clear from (\ref{ABC}) since, for example,
any $A$ particles which penetrate the $B$ domain will be driven backwards
by the dynamics. On the other hand $BA$, $CB$ or $AC$ walls
are all unstable objects and would be quickly eliminated by the dynamics.

In the special case of exactly equal numbers of $A, B$ and $C$
particles it was shown that the model actually obeys detailed balance
with respect to a long range asymmetric, energy function. In fact the
energy is non-extensive in the sense that most configurations have
energies of order $N^2$\cite{EKKM}.  The partition sum was calculated in the
large $N$ limit (with $q$ fixed) and shown to depend linearly on
$N$. This reflects the fact that the phase separation is into three
pure domains and the partition sum is dominated by the $N$ equivalent
translations of the structure comprising three pure domains.  When the
numbers of particles are not identical, detailed balance does not hold
but the phase separation into pure domains remains.  Similar behaviour
has been found in other related models with conserving dynamics
\cite{AHR,LBR}.  Another interesting model is where phase separation
occurs on a quasi-one-dimensional system ($2 \times N$ sites) but not
on a strictly one-dimensional system \cite{KSZ}. It should also be
mentioned that systems with a cyclic symmetry but with non-conserving
dynamics have been studied and shown to order into a frozen
state \cite{FKB}.

Any discussion of nonequilibrium phase transitions is not complete
without mentioning the most well known class, that of directed
percolation. Various models are reviewed elsewhere in this
volume\cite{Haye} so here I just sketch the basic behaviour by
referring to a particular model, the contact process
\cite{Liggett,MD}.  Each site of a lattice is either empty or contains
a particle. Particles are annihilated with rate $1$ and particles are
created at empty sites with rate $n\lambda/2$ where $n$ is the number
of occupied nearest neighbours of the site ($n=0,1,2$). Note that the
`inactive state' where all sites are empty is an absorbing state.
Above a critical value of $\lambda$ there is a finite probability that
starting from a single particle on an infinite lattice, the system
will remain active as $t \to \infty$. This phase transition has
well-studied associated critical exponents and scaling behaviour.
Moreover it appears to be a universality class in the sense
that the same exponents are found in all systems, with the same
symmetry and conservation laws, exhibiting a phase transition from an
absorbing inactive state to an active state\cite{MD}.

However as described so far the contact process is distinct from the
other hopping particle models discussed in that on any finite lattice
the absorbing state is reached in a finite time and is therefore the
steady state. The active state only becomes available as a steady
state on an infinite system.  We mention briefly that it is in fact
possible to define hopping particle models, similar in spirit to the
nonequilibrium models discussed in previous paragraphs, that exhibit
phase transitions connected with directed percolation.  These models
can have non-conserved order parameter \cite{AEHM} or conserved order
parameter \cite{HMS}. Although there are no absorbing states in these
models, they have the common feature of certain microscopic
processes being forbidden.

A final class of transitions in one-dimensional hopping particle
models is that  involving spatial condensation, whereby a finite
fraction of the particles condenses onto the same site. Examples
include the appearance of a large aggregate
in models of aggregation and fragmentation\cite{MKB} and
the emergence of a single flock in dynamical models
of flocking  \cite{CBV,OE}. In the Section~5 we shall examine
a very simple example of a condensation transition which occurs
in the zero-range process and see how it is related to a defect induced
transition.

\section{The zero-range process}

The zero-range process was first introduced into the mathematical
literature as an example of interacting Markov processes
\cite{Spitzer}. Since then the mathematical achievements have been to
prove existence theorems, invariant measures and hydrodynamic limits
\cite{Liggett,Spohn}.

It is not widely appreciated that the zero-range process has many
physical applications; moreover it has often appeared incognito in a
wide range of different contexts.  Examples include  the repton model
of polymer dynamics with periodic boundary conditions \cite{vLK}; a
model of sandpile dynamics \cite{CGS}; the backgammon model
\cite{Ritort} for glassy dynamics due to entropic barriers; the
drop-push model for the dynamics of a fluid moving through backbends
in a porous medium \cite{BR};
microscopic models of step flow growth \cite{KS,KrugChap}
and a bosonic lattice gas
\cite{KKRP}. We shall discuss some of these in the
sequel.  The zero-range process is also closely related to the more widely
known asymmetric exclusion process \cite{Spitzer,Liggett} as we shall
describe below.

\subsection{Model definition}
\label{Sec:ZRP}

In general one can consider the zero-range process on a lattice of
arbitrary dimension, and of (countably) infinite or finite number of
sites. Initially Spitzer \cite{Spitzer}
considered a finite number of sites. However, subsequently
most mathematical works tackle the invariant measure on an
infinite system \cite{Andjel}. 
For our purposes, it is most convenient to
consider a finite system, compute the steady state and only then take
the limit of an infinite system. Note that the steady state $t \to
\infty$ and the infinite volume limit do not necessarily commute
{\it e.g.} on an infinite system the invariant measure (steady state)
is not necessarily unique.

We consider a one-dimensional finite lattice of $M$ sites with sites
labelled $\mu =1 \ldots M$ and periodic boundary conditions. Each site
can hold an integer number of indistinguishable particles.  The
configuration of the system is specified by the occupation numbers
$n_{\mu}$ of each site $\mu$.  The total number of
particles is denoted by $L$ and is conserved under the dynamics.
The dynamics of the system is given by
the rates at which a particle leaves a site $\mu$ (one can think of it
as the topmost particle---see Figure 1a).    As
our first example we assume it moves to the left nearest neighbour
site $\mu{-}1$. The hopping rates $u(n)$ are a function of $n$ the
number of particles at the site of departure.  Some particular cases
are: if $u(n) =n$ then the dynamics of each particle is independent of
the others; if $u(n) = {\rm const}\quad$ for $n>0$ then the rate at
which a particle leaves a site is unaffected by the number of particles
at the site (as long as it is greater than zero).  It is helpful to
think of performing a Monte-Carlo simulation: in the $u(n) =n$ case at
each update a particle would be picked at random and moved to its
nearest neighbour site; in the $u(n) =$ constant case a site would be
picked at random and a single particle moved to the nearest neighbour
site.

A possible source of confusion in the definition of the model is that
in \cite{Spitzer} and some other papers the hop rates $u(n)$ are defined
as the hop rate {\it per} particle at a site; thus $u(n)$ in those
works are $1/n$ of the $u(n)$ defined here.

The important attribute of the zero-range process is that it yields a steady state
described by a product measure. By this it is meant that the steady state
probability $P( \{ n_\mu \})$
of finding the system in configuration 
$\{n_1, n_2 \ldots n_M\}$ is given by a product of factors
$f(n_\mu)$ often referred to as marginals
\begin{equation}
P( \{ n_\mu \}) = \frac{1}{Z(M,L)} \prod_{\mu=1}^{M} f ( n_{\mu} )\;.
\label{Prob}
\end{equation}
Here the normalisation $Z(M,L)$ is introduced so that the sum of the
probabilities for all configurations,
with the correct number of particles $L$, is one.  We shall explore later
in Section~\ref{Sec:cond} the interesting possibilities afforded by the
form (\ref{Prob}).

In the basic model described above, $f(n)$ is given by
\begin{eqnarray}
f(n) &=&  \prod_{m=1}^{n} \frac{1}{u(m)}\quad\mbox{for}\quad n\ge 1
\nonumber \\
  &=& 1  \quad\mbox{for}\quad n=0
\label{f1}
\end{eqnarray} 
Note that $f(n)$ is defined only up to a multiplicative constant
and we could have included a factor
$z^n$ in (\ref{f1}). Later this
factor reappears as a fugacity in Section~\ref{Sec:cond}.

The proof of (\ref{Prob},\ref{f1}) is, happily, straightforward. One
simply considers the stationarity condition on the probability of a
configuration (probability current out of the configuration due to
hops is equal to probability current into the configuration
due to hops):
\begin{equation}
\sum_{\mu} 
\theta (n_\mu) u(n_\mu) P(n_1 \ldots n_\mu \ldots n_L) 
=
\sum_{\mu} \theta (n_\mu)
u(n_{\mu{+}1}{+}1) P(n_1 \ldots n_{\mu}{-}1, n_{\mu{+}1}{+}1 \ldots n_L) \;.
\label{station}
\end{equation}
We have included the Heaviside function to highlight that it is the
sites with $n>1$ that allow exit from the configuration (lhs of
(\ref{station})) but also allow entry to the configuration (rhs of
(\ref{station})).  Equating the terms $\mu$ with $\mu >1$ and
cancelling common factors assuming (\ref{Prob}), results 
(for $n_\mu \geq 1$) in
\begin{equation}
u(n_\mu)f(n_{\mu-1}) f(n_{\mu}) = u(n_{\mu+1}+1) f(n_{\mu}-1) f(n_{\mu+1}+1)
\end{equation}
This equality can be recast as
\begin{equation}
u(n_\mu) \frac{f(n_{\mu})}{f(n_{\mu}-1)}
= u(n_{\mu{+}1}+1)\frac{ f(n_{\mu{+}1}+1)}{ f(n_{\mu{+}1})}
=\mbox{constant}
\end{equation}
Setting the  constant equal to unity implies
\begin{equation}
f(n_\mu) =  \frac{f(n_{\mu}-1)}{ u(n_\mu)}
\label{recurr}
\end{equation}
and iterating (\ref{recurr}) leads to (\ref{f1})
where we have chosen $f(0)=1$.
\begin{figure}[htb]
\begin{center}
\includegraphics[scale=0.66]{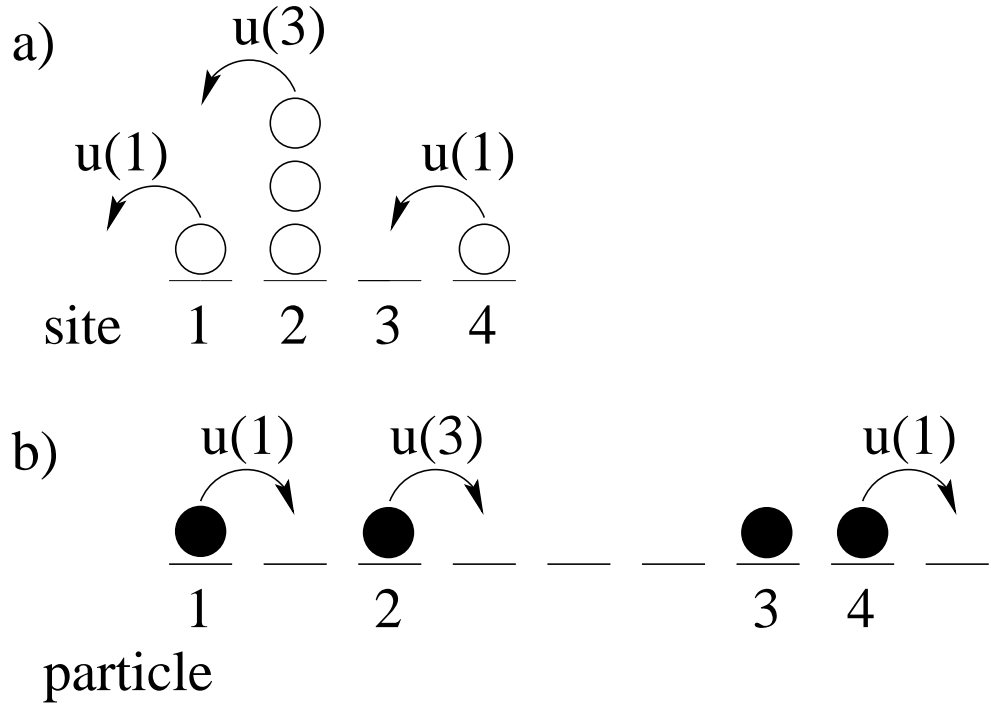}
\end{center}
\caption{\label{fig:ZRP} Equivalence of
zero range process and asymmetric exclusion process.}
\end{figure}

\subsection{Relation to the  asymmetric exclusion process}
\label{Sec:map}
There exists an exact mapping from a zero-range process to an
asymmetric exclusion process. This is illustrated in Figure~1.  The
idea is to consider the particles of the zero-range process as the
holes (empty sites) of the exclusion process.  Then the sites of the
zero-range process become the moving particles of the exclusion
process.  This is possible because of the preservation of the order of
particles under the simple exclusion dynamics. Note  that in the exclusion
process we have $M$ particles hopping on a 
lattice of $M+L$ sites

An interesting feature of the mapping is that it converts a model
where the local degree of freedom can take unbounded values (particle
number in the zero-range process) to a model where the local site
variable is restricted to two values. On the other hand, a hopping
rate $u(m)$ which is dependent on $m$ corresponds to a hopping rate in
the exclusion process which depends on the gap to the particle in
front. So in principle the particles can feel each other's presence
and it is possible to have a long-range interaction.

\section{Generalisations}
We now show how the totally asymmetric, homogeneous zero-range process
we have considered so far may be generalised yet retain steady states
of a similar form to (\ref{Prob},\ref{f1}).

\subsection{Inhomogeneous system}
First we consider an inhomogeneous
system by which we mean the  hopping rates  are site dependent:
the hopping rate out of site $\mu$ when it contains
$n_\mu$ particles is $u_\mu(n_\mu)$.
It is easy to check that the steady state is simply
modified to
\begin{equation}
P( \{ n_\mu \}) = \frac{1}{Z(M,L)} \prod_{\mu=1}^{L} f_{\mu} ( n_{\mu} )\;.
\label{Prob2}
\end{equation}
where $f_\mu$ are given by
\begin{eqnarray}
f_\mu(n) &=& \prod_{m=1}^{n} \frac{1}{u_\mu(m)}
\quad\mbox{for}\quad n\ge 1
\nonumber \\
  &=& 1 \quad\quad \quad\mbox{for}\quad n=0
\label{f2}
\end{eqnarray}
The proof is identical to that given above for the homogeneous
case, with the trivial replacement of $u(n_\mu)$ by $u_\mu(n_\mu)$

\subsection{Discrete Time Dynamics}
A further generalisation is to the case of
discrete time dynamics.
This has been studied in \cite{MRE97} in the context of a disordered
asymmetric exclusion process. Here we translate
the results  into the zero-range process. Rather than processes
occurring with a rate, time is counted in discrete steps
and at each time step events occur with certain probabilities.

In the case of {\it Parallel Dynamics}, at each time-step all sites are
updated. One particle from each site $\mu$ is moved to the
left, each
with probability $p_\mu(n_\mu)$ where $n_\mu$ is the number of
particles at the site before the update.  Note that  the particles
move simultaneously and particles do not move more than one site.

It turns out that the steady state again has the form
(\ref{Prob2}).   It was shown in \cite{MRE97} that
\begin{eqnarray}
f_\mu( n )
 &=& 1- p_\mu(1)\;\;\;\mbox{for}\;\;\; n =0 \nonumber \\
 &=&  \frac{1-p_\mu(1)}{1-p_\mu(n)}
\prod_{m=1}^{n}\frac{1-p_\mu(m)}{p_\mu(m)}
\;\;\;\mbox{for}\;\;\; n >0 \;.
\label{fPgen}
\end{eqnarray}

To recover the continuous time dynamics we can call the interval between
time steps $dt$ and let 
$p_\mu(n_\mu) = u_\mu(n_\mu) dt$. Then continuous time
dynamics is given by the limit $dt \to 0$ and, to
within a constant factor $dt^n$,
(\ref{fPgen}) recovers (\ref{f2}).
In this way one can interpolate between discrete time,
parallel dynamics and continuous time dynamics.

In \cite{MRE97}  ordered sequential updating schemes were
also  considered.
These are discrete time updating schemes were one site is updated
at a time, but the sites are updated in a fixed order.
The steady states for the forwards and backwards updating sequences
were derived and it turns out they too have the form 
(\ref{Prob2}) with $f_\mu$ taking an expression related
to the parallel case (\ref{fPgen}).

\subsection{Arbitrary Network}
\label{Sec:Net}
In the original paper paper of Spitzer\cite{Spitzer} some more general
versions of the zero range process were considered.  Here we discuss
one interesting case which serves to generalise the (totally
asymmetric) zero range process defined above to a process on  a more
general lattice or for any finite collection of points with
a prescribed transition matrix for the dynamics of a single particle
\cite{Andjel}. 

In this case the rate of hopping of a particle at site $\mu$
containing $n_\mu$ particles is equal to $u_\mu(n_\mu)$ and the probability
that a particle leaving site $\mu$ will move to site $\nu$ is denoted
$W(\mu{\to}\nu)$.  Thus the probability that in time $dt$ a
particle at $\mu$ moves to $\nu$ is
\begin{equation}
 u_\mu(n_\mu) W(\mu{\to}\nu)\ dt\;.
\label{hop}
\end{equation}
Note that the probabilities
$W(\mu{\to}\nu)$ define a stochastic matrix for a single
particle moving on a finite collection of $M$ sites and we take
\begin{equation}
\sum_{\nu} W(\mu{\to}\nu) = 1\; .
\end{equation}
so that probability is conserved.
We refer to the collection of points
together with the
prescribed transition matrix 
$W(\mu{\to}\nu)$ as the network.

We assume that the transition matrix is irreducible
({\it i.e.} the particle can pass from any given point
to any other after sufficient time and the system is ergodic) 
so that we have a unique
steady state probability for the single particle problem:
\begin{equation}
s_\nu = \sum_{\mu} s_\mu W(\mu{\to}\nu) \;.
\label{spss}
\end{equation}

We now show that the steady state  for the many-particle problem
defined above is given by (\ref{Prob2})
where now $f_\mu(n)$ is given by
\begin{eqnarray}
f_\mu(n) &=& \prod_{m=1}^{n}\left[ \frac{s_\mu}{
 u_\mu(m)}\right]
\quad\mbox{for}\quad n\ge 1
\nonumber \\
  &=& 1 \quad\quad \quad\mbox{for}\quad n=0
\label{f3}
\end{eqnarray}
The proof is again a straightforward generalisation of that of 
Section~\ref{Sec:ZRP}. Equation
(\ref{station}) is modified to
\begin{equation}
\sum_{\mu} 
\theta (n_\mu) u_\mu(n_\mu) P(n_1 \ldots  n_L) 
=
\sum_{\mu} \sum_{\nu{\neq}\mu}
\theta (n_\mu) p(\nu{\to}\mu)
u_\nu(n_{\nu}{+}1) P(n_1 \ldots n_{\nu}{+}1\ldots n_{\mu}{-}1 \ldots n_L) \;.
\label{station3}
\end{equation}
Equating the terms $\mu$ on each side of (\ref{station3}),
assuming (\ref{Prob2}) and cancelling common factors
yields
\begin{equation}
\theta (n_\mu) u_\mu(n_\mu)f_\mu(n_\mu)
=
\theta (n_\mu) \sum_{\nu{\neq}\mu}
 W(\nu{\to}\mu)
u_\nu(n_{\nu}{+}1)
 f_\mu(n_\mu{-}1) \frac{f_\nu(n_\nu{+}1)}{  f_\nu(n_\nu) } \;.
\end{equation}
Inserting (\ref{f3}) leads to the condition
\begin{equation}
s_\mu = \sum_{\nu\neq\mu} s_\nu W(\nu{\to}\mu) 
\label{spss2}
\end{equation}
which is the same as the
single particle steady state condition (\ref{spss}).

A simple case considered by Spitzer is when $W(\mu{\to}\nu)$ is a
doubly stochastic matrix which is defined by the property
\begin{equation}
\sum_\nu W(\nu{\to}\mu)= \sum_\mu W(\nu{\to}\mu)= 1\;.
\label{dstoch}
\end{equation}
Equations (\ref{dstoch}) and (\ref{spss}) then imply
that the single particle problem has a homogeneous
steady state  $s_\mu=$ constant.

Let us also discuss an example where the single particle
problem has an inhomogeneous steady state.
We consider a one-dimensional lattice
where hops
to the left and right
neighbours are allowed but with probabilities that depend
on the site. Thus, we may write
\begin{eqnarray}
W(\mu{\to}\nu) &=& q_\mu\quad\mbox{for}\quad\nu =\mu+1\\
&=& 1-q_\mu\quad\mbox{for}\quad\nu =\mu-1\\
&=& 0\quad \mbox{otherwise}\;.
\end{eqnarray}
The steady state of 
the single particle problem (random walker on 
a disordered one dimensional lattice \cite{Derrida}) 
\begin{equation}
s_\mu= (1-q_{\mu+1}) s_{\mu+1} + q_{\mu-1} s_{\mu-1}
\end{equation}
can be  solved
and one obtains
\begin{equation}
 s_{\mu} =\left[
                         \sum_{i=0}^{M-1} \frac{1}{q_{\mu-i}}
                           \prod_{\nu = 0}^{i}
         \frac{q_{\mu-\nu}}{1-q_{\mu -\nu}}
                           \right]\;.
\label{rw}
\end{equation}

This network is relevant in the disordered one-dimensional exclusion
process studied in \cite{BFL,MRE96,KF}.  The sites in the present
model correspond to the particles in the exclusion process which each
have their own forward and backward hopping rates.  Another,
particular instance of this network occurs in \cite{vLK}, where a
repton model of gel electrophoresis is studied in the case of periodic
boundary conditions (see Section~\ref{Sec:app}).

In special cases of the zero-range process detailed balance may hold.
The condition for this is
\begin{equation}
u_\mu(n_\mu) W(\mu{\to}\nu) P(n_1 \ldots  n_L)
=
u_\nu(n_\nu+1) W(\nu{\to}\mu) P(n_1 \ldots n_\nu{+}1 \ldots n_\mu{-}1\ldots n_L)\;.
\end{equation}
Substituting the form (\ref{Prob2},\ref{f3}) leads to the condition
\begin{equation}
s_\mu W(\mu{\to}\nu)
=
s_\nu W(\nu{\to}\mu)
\label{DB}
\end{equation}
which is just the detailed balance condition for the single particle
problem.

An interesting consequence of the form of the steady state (\ref{f3})
is that it allows one to relate an arbitrary zero-range process to a
model obeying detailed balance. The idea is that if detailed balance
doesn't hold, we can always define a new zero-range process (to be
denoted by a prime) with the same steady state, but with a different
dynamics obeying detailed balance.  To do this, we solve the single
particle problem (\ref{spss}) for the original model to obtain
$s_\mu$.  For any collection of points we can always define a new
single particle transition matrix $W'(\mu{\to}\nu)$ that satisfies
detailed balance with respect to a homogeneous steady state ($s'_\mu$
= constant).  The new model is defined by a new set of hopping rates
$u'_\mu(m) =u_\mu(m)s_\mu'/s_\mu$ together with the new transition
matrix $W'(\mu{\to}\nu)$.  It is easy to check from (\ref{f3}) that
the new model has the same steady state as the original.

Thus, within the realm of zero-range processes, to the steady state of
any nonequilibrium model we can always identify a model satisfying
detailed balance and therefore an energy function.  Of course,
although the steady states are the same, there is no reason for the
dynamical behaviour of the two systems to be related.  To clarify this
point we will discuss a simple example in Section~\ref{Sec:Condinhom}.

The marginals (\ref{f3}) have the interesting structure of being a
product of a term $(s_\mu)^n$ that depends on the nature of the
network and a term involving the product of $u_\nu(m)$ which reflects
the interactions at the site.  The network can represent an arbitrary
dimensional lattice or the effects of disorder, the only difficulty to
surmount in obtaining the steady state is the solution of the single
particle problem.

\section{Condensation Transitions}
\label{Sec:cond}
We now proceed to analyse the steady states of form (\ref{Prob2}) and
the condensation transition that may occur. The important quantity to
consider is the normalisation $Z(M,L)$ as it plays the role of the
partition sum.  The normalisation is defined through the condition
\begin{equation}
Z(M,L) = \sum_{n_1,n_2 \ldots n_M}
\delta(\sum_{\mu} n_\mu{-}L)
 \prod_{\mu=1}^{M} f_\mu( n_\mu )
\label{norm}
\end{equation}
where the $\delta$ function enforces the constraint of $L$ particles.  The
normalisation may be considered as the analogue of a 
canonical partition
function of a thermodynamic system.

We define the `speed' $v$ as the average
hopping rate out of a site
\begin{eqnarray}
v&=& \frac{1}{Z(M,L)}
 \sum_{n_1,n_2 \ldots n_M}
\delta(  \sum_{\mu} n_\mu{-}L)
u(n_1)
 \prod_{\mu=1}^{M} f_\mu( n_\mu )
\nonumber \\
&=&\frac{Z(M,L{-}1)}{Z(M,L)}
\label{speed}
\end{eqnarray}
where we have used (\ref{Prob2},\ref{f2}).  Note that (\ref{speed})
tells us that the speed is independent of site and thus may be
considered a conserved quantity in the steady state of the system.  In
the totally asymmetric system considered in Section~\ref{Sec:ZRP} the
speed is equal to the current of particles flowing between
neighbouring sites and is clearly a conserved quantity in the steady
state.  More generally, however, the speed is not equal to the current
and the fact that the speed is a conserved quantity is not a priori
obvious.  For example, in a system obeying detailed balance the (net)
current is zero, but the speed as defined above remains finite.  The
speed is a ratio of partition functions of different system sizes
(\ref{speed}) and corresponds to a fugacity, as we shall see below.

We will consider also the probability distribution of the number of
particles at a site, taken here to be site $1$
\begin{eqnarray}
P_1(n_1=x) &=& \frac{1}{Z(M,L)}
 \sum_{n_2 \ldots n_M}
\delta( x+n_2+ \ldots n_M{-}L)
f_1(x) \prod_{\mu=2}^{M} f_\mu( n_\mu )
\nonumber \\
&=&f_1(x) \frac{Z(M{-}1,L{-}x)}{Z(M,L)}
\label{gapdist}
\end{eqnarray}
(where $Z(M{-}1,L{-}x)$ is the partition function for a system with
site 1 removed). In general the probability distribution is site
dependent but for a homogeneous system ($f_\mu$ independent of $\mu$)
it will be the same for all sites.

We now use the integral representation of the delta function to write
the partition function as
\begin{equation}
Z(M,L) = \oint \, \frac{dz}{2\pi i} \ z^{-(L+1)}\ 
\prod_{\mu=1}^{M} F_\mu(z)\; ,
\label{Zint}
\end{equation}
where 
\begin{equation}
F_\mu(z) = 
 \sum_{m=0}^{\infty} z^{m} \ f_\mu(m)\; .
\label{Fdef}
\end{equation}
For large $M,L$ (\ref{Zint}) is dominated by the saddle point of the
integral and the value of $z$ at the saddle point is the
fugacity. The equation for the saddle point reduces to
\begin{equation}
\frac{L}{M} = \frac{z}{M} \sum_{\mu=1}^{M} \frac{\partial}{\partial z}
\ln F_\mu(z) \label{sad}
\end{equation}
which, defining $\phi= L/M$, can be written as
\begin{equation}
\phi =  \frac{z}{M} \sum_{\mu=1}^{M}  \frac{F'_\mu(z)}{F_\mu(z)}\;.
\label{sad2}
\end{equation}
In the thermodynamic limit,
\begin{equation}
M \rightarrow \infty \;\;\;\mbox{with}\;\;\; L =\phi M\; ,
\label{thermlim}
\end{equation}
where   the density $\phi$ is held fixed,
the question is whether a valid saddle point value
of z can be found from
(\ref{sad2}). We expect that for low $\phi$  the saddle point
is valid but,
as we shall discuss, there exists a maximum value of $z$ 
and if at this maximum value the rhs of (\ref{sad2}) is finite,
then for large $\phi$ (\ref{sad2}) cannot be satisfied.
We now consider separately, and in more detail, how condensation may occur in
the inhomogeneous
and the homogeneous case.

\subsection{Inhomogeneous case}
\label{Sec:Condinhom}
In general, the inhomogeneous case {\it i.e.}  where $F_\mu(z)$
depends on the site $\mu$ through (\ref{f3}), is difficult to analyse.
Here we would just like to give an idea of how a condensation
transition may occur by discussing a simple example. We then go on to
analyse perhaps the simplest example of a condensation transition: a
single inhomogeneous site \cite{MRE96}.

First we take the general model discussed in Section~4.3 and set
$u_\mu(m) = u_\mu$ for $m>0$ {\it i.e.} the hopping rate does not
depend on the number of particles at a site. We consider doubly
stochastic transition matrices $W(\mu{\to}\nu)$ (see Eq. \ref{dstoch})
 so that we may take
$s(\mu) =constant$ and without loss of generality we set the constant
equal to one.  For the moment we do not specify further the transition
matrix; later we will discuss two specific examples one obeying
detailed balance and one not.  Under these conditions $f_\mu$ is given
by
\begin{equation}
f_\mu(n) = \left( \frac{1}{u_\mu}\right)^{\! n_\mu}
\end{equation}
and the probability of occupancies $\{n_1,n_2,\ldots,n_M\}$ is
\begin{equation}
P(\{ n_1,n_2,\ldots,n_M\}) = \frac{1}{Z(M,L)}
\prod_{\mu =1}^{M}
\left(\frac{1}{u_\mu}\right)^{\! n_\mu}\;.
\end{equation}
The mapping to an ideal Bose gas is evident: the $L$ particles of the
zero-range process are viewed as Bosons which may reside in $M$ states
with energies $E_{\mu}$ determined by the site hopping rates:
$\exp(-\beta E_{\mu}) = 1/u_{\mu}$.  Thus the ground state corresponds
to the site with the lowest hopping rate.  The normalisation $Z(M,L)$
is equivalent to the canonical partition function of the Bose gas.  We
can sum the geometric series (\ref{Fdef}) to obtain $F_\mu$ and
$F'_\mu$ then taking the large $M$ limit allows the sum over $\mu$ to
be written as an integral
\begin{equation}
\phi = \int_{u_{\rm min}}^{\infty} du  {\cal P}(u) \  \frac{z}{u-z}
\label{gce2}
\end{equation}
where ${\cal P}(u)$ is the probability distribution of site hopping rates
with $u_{\rm min}$ the lowest possible site hopping rate.  Interpreting
${\cal P}(u)$ as a density of states, equation (\ref{gce2})
corresponds to the condition that in the grand canonical ensemble of
an ideal Bose gas the number Bosons per state is $\phi$.  The theory
of Bose condensation \cite{Huang} tells us that when certain
conditions on the density of low energy states pertain we can have a
condensation transition.  Then (\ref{sad2}) can no longer be satisfied
and we have a condensation of particles into the ground state, which
is here the site with the slowest hopping rate.  This case is discussed
further, in the context of an asymmetric
exclusion process on an infinite system, by
J. Krug in this volume \cite{Krug99}.

We now turn to  the simplest case of an inhomogeneous system:
site 1 has  $u_1=p$ while the other $M-1$ particles have hopping rates 
$u_{\mu}=1$ when $ \mu >1$. It is easy to see that (\ref{Prob2})
simplifies to
\begin{equation}
P(\{ n_\mu \})= \frac{1}{Z(M,L)}\frac{1}{p^{n_1}}
\label{Probdefect}
\end{equation}
In this case 
the normalisation $Z(M,L)$ is easy to calculate combinatorially:
\begin{eqnarray}
 Z(M,L) &=&  \sum_{n_1,n_2 \ldots n_M}
\delta(  \sum_{\mu} n_\mu{-}L)
  p^{-n_1}
\nonumber \\
&=& \sum_{n_1=0}^{L}
\left( \begin{array}{c} L{+}M{-}n_1{-}2 \\ M{-}2 \end{array} \right) p^{-n_1} \; ,
\label{z1slow}
\end{eqnarray}
yielding an exact expression for the speed through
(\ref{speed}).
In the thermodynamic limit 
the sum (\ref{z1slow}) is dominated by 
$n_1 \sim
{\cal O}(1)$ for $\phi < p/(1-p)$ and $n_1 \sim {\cal O}(L)$ for $\phi > p/(1-p)$ and it can be shown that
\begin{eqnarray}
\mbox{for}\;\;\; \phi< \frac{p}{1-p} \;\;\; &Z(M,L) \simeq
\left( \begin{array}{c} L+M \\ M \end{array} \right)
\frac{1}{1+\phi}\ \frac{p}{p-\phi(1-p)}&
\label{ldensity}
 \;\; \mbox{and}\;\; v \rightarrow 1-\rho \\
\mbox{for}\;\;\; \phi > \frac{p}{1-p} \;\;\; &Z(M,L) \simeq  p^{-L}
 \ (1-p)^{-(M-1)}&
 \;\; \mbox{and}\;\; v \rightarrow \phi/(1+\phi)
\label{hdensity}
\end{eqnarray}
In the high density phase, defined by (\ref{hdensity})
we have a condensate since the average
number of particles at site 1 is $\langle n_1 \rangle/L = \phi -p/(1-p)$.
In the low density phase (\ref{ldensity}) the particles are evenly spread
between all sites and we will refer to it as the homogeneous phase.

We now discuss two models which both have this steady state:
a driven system and a system obeying detailed balance.
This provides an illustration of the idea discussed
in Section~4.3 whereby a zero range process not obeying detailed
can be related to one obeying detailed balance.

First we take the totally asymmetric model so that particles 
move to the site to the left:
the transition matrix is 
$$W(\mu{\to}\nu) = \delta_{\nu,\mu{-}1}\;.$$
So this model is
similar to that discussed in Section~\ref{Sec:ZRP}, and a mapping to a
totally asymmetric exclusion process can be made in the same way as
Section~\ref{Sec:map}.  The equivalent exclusion process is
illustrated in Figure~2.  We see that the equivalent
exclusion process is system of hard-core
particles hopping to the right, one particle being slower than the
rest.  The interpretation of the two phases within the context of the
exclusion process is that in the condensed phase (for the exclusion
process a low density of particles) a `traffic jam' forms behind the
slow particle and the slow particle has a finite fraction of the
lattice as `empty road' ahead. Whereas in the homogeneous phase (a high
density of particle for the exclusion process) the particles are
roughly evenly spaced.
\begin{figure}[htb]
\begin{center}
\includegraphics[scale=0.66]{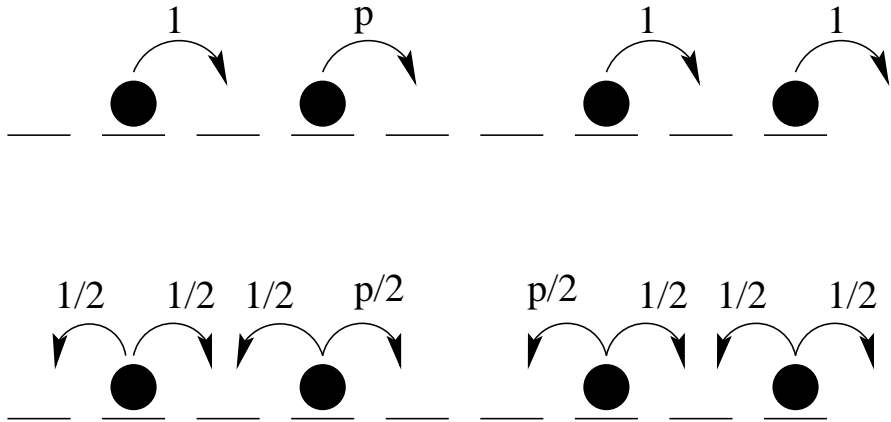}
\end{center}
\label{fig:equiv} 
\caption{
A totally asymmetric exclusion model (upper) and a model with zero current
(lower) that have equivalent steady states leading to (\ref{z1slow}).}
\end{figure}

On the other hand one may consider the case
where the one particle problem is a symmetric random walk 
so that the system
obeys detailed balance. The transition matrix is given by
$$W(\mu{\to}\nu) = 
\frac{1}{2}\delta_{\nu,\mu{-}1} +\frac{1}{2}\delta_{\nu,\mu{+}1}\;.$$
When we map this system to a simple exclusion
process we see from Figure~2 that we have a system of
particles, the bulk of which perform a symmetric exclusion dynamics
but with two adjacent asymmetric particles: the left one biased to the
left and the right one biased to the right. 
In the condensed phase
the gap between these particles diverges.  
Previously a single asymmetric particle in a sea of symmetric
particles has been studied \cite{BOMR}
but in that case there is no phase transition.
At first it seems that we
have found a counterexample to the received wisdom that no phase
transition should occur in an equilibrium system, since we have a
condensation transition in a model with local dynamics obeying
detailed balance.  Inferring an energy function from the
steady state  (\ref{Probdefect}) through the following equation
$$  \exp\left[-(\beta E)\right] 
=  \exp\left[ - (x_2 -x_1) \ln p \right]\;$$
reveals that our effective energy increases linearly with distance
$x_2-x_1$ between the two asymmetric particles. Therefore
the energy is `unphysical' in that it has very long range interactions.
Thus the phase transition can be rationalised
within  the categories of exceptions
discussed in Section~\ref{Sec:equil}

We have seen that this simplest example
of a condensation transition (a single
inhomogeneous site in the zero range process) is exhibited both in a
driven model and also in a model obeying detailed balance but with
long-range energy function. Again it should be stressed that although
the steady states of these two models are equivalent, the dynamical
properties should be very different. For example in the homogeneous
phase of the driven model we expect asymmetric exclusion like
behaviour and the dynamic exponent should be $3/2$ implying relaxation
times of $M^{3/2}$ on a finite system. However in the homogeneous
phase of the model obeying detailed balance we expect symmetric
exclusion like behaviour and the dynamic exponent to be $2$ implying
relaxation times of $M^2$ \cite{GS}.

\subsection{Homogeneous case}
\label{Sec:condhom}
We now consider the homogeneous zero-range process where in
(\ref{hop}) the hopping rates are site independent and the single
particle problem (\ref{spss}) has a homogeneous steady state $s_\mu=1$
\cite{OJO}.  A similar analysis has been carried out in the context of
balls-in-boxes and branched polymer models \cite{BBJ}.

In the present case, (\ref{Fdef}) is independent of $\mu$ and reads
\begin{equation}
F(z) = \sum_{n=0}^\infty   \prod_{m=1}^n \left[ \frac{z}{u(m)} \right]
\label{Fhom}
\end{equation}
The fugacity $z$ must be chosen so that $F$ converges or else we could
not have performed (\ref{Fdef}).  Therefore $z$ is restricted
to $z \leq \beta$ where we
define $\beta$ to be the radius of convergence of $F(z)$.
From (\ref{Fhom}) we see that
$\beta$ is the limiting value value of the
$u(m)$ {\it i.e.} the limiting value
of the  hopping rate out of a site
for a large number of particles at a site.  We interpret
(\ref{sad2}) as giving a relation between the density of holes (number
of holes per site) and the fugacity $z$. 
The saddle point condition (\ref{sad2}) becomes 
\begin{equation}
\label{sad3}
\phi =   \frac{z F'(z)}{F(z)}
\end{equation}
Given that the rhs of (\ref{sad3}) is a monotonically increasing function
of $z$ (which is not difficult to prove) we deduce that density of
particle increases with fugacity.  However if at $z=\beta$, the
maximum allowed value of $z$, the rhs of (\ref{sad3}) is still finite
then one can no longer solve for the density and one must have a
condensation transition. Physically, the condensation would correspond
to a spontaneous symmetry breaking where one of the sites is
spontaneously selected to hold a finite fraction of the particles.

Thus, for condensation to occur ({\it i.e.} when $\phi$ is large enough
for (\ref{sad3}) not to have a solution for the allowed values of $z$) we
require
\begin{equation}
\lim_{z\to \beta} \frac{F'(z)}{F(z)} < \infty\; .
\end{equation}
We now assume that $u(n)$ decreases uniformly to $\beta$ in the large $n$
limit as
\begin{equation}
u(n) = \beta( 1 + \zeta(n) )
\end{equation}
where $\zeta(n)$ is a monotonically decreasing function.
Analysis of the series 
\begin{eqnarray}
F(\beta) &=& \sum_{n=0}^{\infty}
 \exp \left\{ - \sum_{m=1}^n \ln\left[1+\zeta(m)\right] \right\}
\nonumber \\
F'(\beta) &=& \sum_{n=0}^{\infty} n
 \exp \left\{ - \sum_{m=1}^n \ln\left[1+\zeta(m)\right] \right\}
\label{series}
\end{eqnarray}
reveals that the condition for condensation is simply that
$F'(\beta)$ is finite and this occurs if $u(n)$ decays to $\beta$ more
slowly than $\beta(1+2/n)$.
  (This is easiest to see by expanding $\ln
\left[1+\zeta\right]$ and approximating the sum over $m$ by an
integral in (\ref{series}).)

In order to fit this result  into the picture of section
Section~\ref{Sec:equil} one can argue that since the condensate has an
extensive number of particles at a site, the local site variable is
unbounded. Therefore the `no phase transition rule' does not apply. One
also gains insight by translating the results into the language of the
simple exclusion process.  In this context we can have condensation if
the hop rate of a particle into a gap of size $n$ decays as
$\beta(1+2/n)$ therefore there is an effective long range interaction.

\subsection{Sharp crossover phenomena}
\label{Sec:cross}
Having  discussed the case where a true phase transition occurs
we now consider a homogeneous example
where, although there is no strict condensation transition,
some interesting crossover phenomena occur \cite{OJO}. 

Consider
\begin{eqnarray}
 u(n)&=&1\quad\mbox{for}\quad n<r \\
 u(n)&=&\beta\quad\mbox{for}\quad n\geq r \;.
\end{eqnarray}
One can interpret these hop rates
as meaning that  a site only distinguishes
whether it contains greater than  $r$ particles.
When we use the
 mapping of section~3.2 to a totally asymmetric exclusion process
$r$ becomes the range of the interaction in the sense that
it is the number of sites ahead upto which
a particle in the exclusion process distinguishes.

When these hopping rates are inserted in (\ref{Fhom}) one obtains
\begin{equation}
F(z) = \sum_{n{=}0}^{r{-}1}z^n + \sum_{n=r}^{\infty}
z^n\beta^{r{-}n{-}1}\;.
\end{equation}
Performing the geometric series  readily yields
\begin{eqnarray}
F(z)&=& \frac{1-z^r}{1-z} + \frac{z^r}{\beta-z} \\
zF'(z) &=&z \left[ \frac{1-z^r}{(1-z)^2} 
- \frac{rz^{r{-}1}}{1-z}+ \frac{z^r}{(\beta-z)^2} 
+\frac{rz^{r{-}1}}{(\beta-z)} \right] \;.
\end{eqnarray}
Then we find the condition (\ref{sad3}) can be written after
a little algebra as
\begin{eqnarray}
(\beta-z)^2 \left[ \phi- z(1+\phi) \right]
&=& z^r(1-\beta)\left[(1{+}\beta{-}2z)z-\phi(1{-}z)(\beta{-}z)\right]
\nonumber \\
&& + rz^r(1-z)(\beta-z)(1-\beta)
\label{solution}
\end{eqnarray}
Therefore for large $r$  we find the solutions
\begin{eqnarray}
\mbox{for}\;\;\; \phi< \frac{\beta}{1-\beta} \;\;\; &z \simeq
\frac{\phi}{1+\phi}- r \left(\frac{\phi}{1+\phi}\right)^r 
\frac{1-\beta}{(1+\phi)\ (\phi-\beta(1+\phi))}
\label{rlow}\\
\mbox{for}\;\;\; \phi= \frac{\beta}{1-\beta} \;\;\; &z \simeq
\beta- \beta^{(1+r)/3} (1-\beta)\\
\mbox{for}\;\;\; \phi> \frac{\beta}{1-\beta} \;\;\; &z \simeq
\beta-  \beta^{(1+r)/2}\frac{1-\beta}{(\phi-\beta(1+\phi))^{1/2}}
\label{rhigh}
\end{eqnarray}
Thus we see as $r{\to}\infty$ we have two phases: a high density phase
$\phi >\beta/(1-\beta)$ where the speed is $\beta$ and a low density
phase where the speed is $\phi/(1+\phi)$. In fact these phases
correspond exactly to those of the single defect problem discussed in
the previous subsection (\ref{ldensity}, \ref{hdensity}) with $\beta$
playing the role of $p$. For finite $r$, $z$ is actually a smoothly
increasing function of $\phi$ but we see from (\ref{rlow},\ref{rhigh})
that the curve  sharpens as $r$ increases. This is illustrated in
Figure~3 where the numerical solution to (\ref{solution}) is plotted
is plotted for $r=10,20,30$. One  sees a dramatic sharpening
as   $r$ increases leading to a sharp crossover between a low density
and high density regime.
\begin{figure}[htb]
\begin{center}
\includegraphics[scale=0.66]{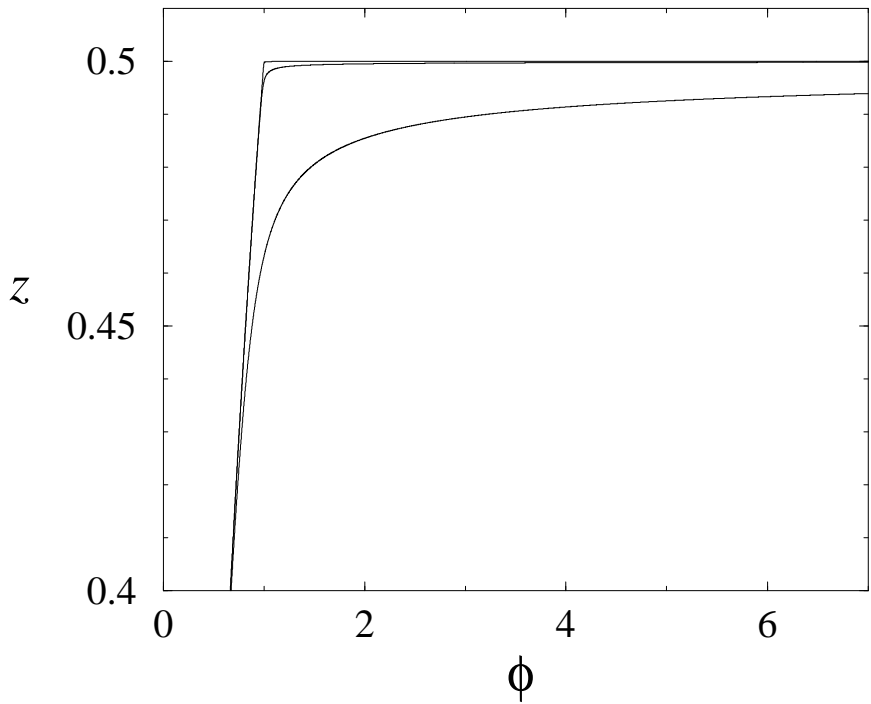}
\end{center}
\label{fig:Sharp}
\caption{ Solutions to
(\ref{solution}) for $\beta= 0.5$ and $r=10,20,30$ (increasing in sharpness of curve).}
\end{figure}

In order to see the effects of this sharp crossover
it is interesting to consider the
particle number  probability distribution (\ref{gapdist}) which for this
system is site independent and given by
\begin{eqnarray}
P(x) &\sim& z^x\quad\mbox{for}\quad x<r \\
 &\sim& (z/\beta)^x\quad\mbox{for}\quad x\geq r \;.
\end{eqnarray}
One can think of this as a sum of two distributions, one
for poorly occupied sites $x<r$ and one
for well occupied sites $x\geq r$.
When $\phi >\beta/(1-\beta)$
the probability distribution for large $x  >r$ goes as
\begin{eqnarray}
P(x)\sim \exp\left\{ - x \beta^{(1+r)/2}\frac{1-\beta}{(\phi-\beta(1+\phi))^{1/2}}\right\}
\end{eqnarray}
so that the typical occupancy of well-occupied sites
goes as $\beta^{-r/2}$. Taking, for example $\beta=0.1$ and
$r=10$ leads to a typical occupancy  of $\sim 10^5$. Therefore to simulate
the model one requires a number of particles very much larger than this!
If care is not taken to do this, and the total
number of particles in the system is comparable 
to the typical occupancy, one would have an apparent condensate
on a finite system.

An example of this phenomenon was studied recently within the context
of a `bus route model' \cite{OEC}.  There the underlying motivation
was to consider how a non-conserved quantity could mediate an
effective long-range interaction amongst a conserved quantity in a
driven system with a strictly local dynamical rule.  The model
considered was defined on a $1d$ lattice. Each site (bus-stop) is
either empty, contains a bus (a conserved particle) or contains a
passenger (non-conserved quantity). The dynamical processes are that
passengers arrives at an empty site with rate $\lambda$; a bus moves
forward to the next stop with rate 1 if that stop is empty; if the
next stop contains passengers the bus moves forward with rate $\beta$
and removes the passengers.  Since the buses are conserved, there is a
well defined steady state average speed $v$. This fact can be used to
integrate out the non-conserved quantity (passengers) within a
mean-field approximation. The idea is that a bus stop, next to bus 1
say, will last have been visited by a bus (bus 2) a mean time ago of
$n/v$ where $n= x_2-x_1$ is the distance between bus 2 and bus 1.
Therefore the mean-field probability that the site next to bus 1 is
not occupied by a passenger is $\exp(-\lambda n/v)$. From this
probability an effective hopping rate for a bus into a gap of size $n$
is obtained by averaging the two possible hop rates $1,\beta$:
\begin{equation}
u(n)=\beta+(1-\beta) \exp(-\lambda n/v)\;.
\end{equation}
We can now see that this mean-field approximation to the
bus-route model is equivalent to a homogeneous zero-range process as discussed
earlier in this section.
Since $u(n)$ decays exponentially,
with decay length  $r=v/\lambda$,  the condition for a strict phase
transition is not met. 
It is reasonable to believe that the system behaves in a similar way
to the system with a finite  `range'  $r$ discussed in Section~\ref{Sec:cross}.
Since $r$ can be made arbitrarily large as $\lambda{\to}0$,
on any finite system an apparent condensation will be seen. In the bus route
problem this corresponds to the universally irritating
situation of all the buses on the route arriving at once.

\section{Some further applications}
\label{Sec:app}
As mentioned earlier the zero-range process and related models have
appeared several times in the modelling of nonequilibrium
phenomena. Here we briefly discuss a few of these instances to
illustrate the ubiquity of the basic model.

In \cite{CGS} models of sandpile dynamics are considered.  A zero
range process is used to model the toppling of sand on a
one-dimensional lattice; specifically the system is homogeneous and
the occupation number of a site becomes the height of sand ($h$) at
that site. The hopping rates are set as $u(1) = 1$ and $u(h)= \lambda$
for $h> 1$, with the transition matrix a symmetric random walk, and
the limit of large $\lambda$ considered. This limit means that a
particle (grain of sand) keeps moving until it finds an unoccupied
site, thus a hopping event may play the role of an avalanche.
(Although in terms of sandpiles and self-organised
criticality this model is rather trivial, it did serve
to investigate the idea of a diverging diffusion constant.)
Note that a slightly different $\lambda\to \infty$ limit (where the
direction of the initial move of the particle is maintained until it
finds an unoccupied site) was also considered but the product measure
is still retained.

In a different context Barma and Ramaswamy \cite{BR} introduced the
`drop-push' model of activated flow involving transport through a
series of traps.  Each trap can only hold a finite number of
particles.  For the trap depth set equal to one this model is
essentially the same as the sandpile model of \cite{CGS} discussed
above ({\it i.e.} it is a zero-range process with some infinite
rates). In fact the version studied in \cite{SRB} is precisely the
limit of $u(n)\to\infty$ for $n>1$ of the totally asymmetric
zero-range process described in Section 3.1.  A generalisation to
inhomogeneous traps, and partially asymmetric hopping rates dependent
on the occupancy of the trap was made in \cite{TB} and a steady state
similar to (\ref{Prob2},\ref{f3}) demonstrated.

The zero-range process is also relevant in the context of $1{+}1$
dimensional interface growth by the step flow mechanism. The interface
can be visualised as an ascending staircase of terraces. Adatoms land
on the terraces and diffuse until they bind to the ascending step. If
the ratio of deposition rates over diffusion rates tends to zero then
the resulting dynamics is that a terrace shrinks by one unit (and the
adjacent higher terrace grows by one unit) with a rate proportional to
the size of the terrace. Thus the terrace lengths are equivalent to
the site occupancies of an asymmetric zero-range process
that was discussed in Section~3.1.  The equivalence of zero
range processes to a general class of step flow models is discussed in
\cite{KrugChap}.

Finally we note that the repton model of gel-electrophoresis
\cite{RubDuke} studied in the case of periodic boundary conditions by
\cite{vLK} is equivalent to an inhomogeneous zero-range process.  In
this case, the particles of the zero-range process represent the
excess stored length of a polymer which diffuses along the tube of the
polymer. The sites in the zero-range process represent the segments of
the polymer tube and the inhomogeneities in site hopping rates
reflect the shape of the polymer tube.

\section{Conclusion}
\label{Sec:Conc}
In this work the aims were to give an overview of the area of phase
transitions and ordering in one-dimensional systems and also to
analyse in some detail a particularly simple model, the zero-range
process.  In section 2 several features were identified which could
lead to the anomalous behaviour of ordering and phase transitions
in equilibrium systems: long-range interactions; zero temperature;
unbounded local variable.  For nonequilibrium systems some concepts
which may be important emerged: conserved order parameter; drive;
forbidden microscopic transitions.

The simplicity of the zero-range process allowed us to analyse the
steady state of the model in detail. First we derived the steady state
for a general class of zero-range processes in Sections 3 and 4. We
then analysed the condensation transitions that can occur. On an
inhomogeneous system the condensation is very reminiscent of
Bose-Einstein condensation. For it to occur requires certain
conditions to hold on the distribution of hopping rates.  In the
homogeneous system the condensation corresponds to a spontaneous
symmetry breaking, since an arbitrary site is selected to hold the
condensate. The condition for it to occur is that the hopping rate
dependence on the site occupancy decays sufficiently slowly.  It was
also shown that when the condition for condensation does not hold, one
can still observe very sharp crossover behaviour and apparent
condensation on a finite system

An interesting possibility that was explored was that of the existence
of an effective energy function.  We saw that any steady state of the
form (\ref{Prob2},\ref{f3}) can be obtained from a process obeying
detailed balance. However when the effective energy is inferred for
cases where phase transition occurs (as was carried out for an
explicit example in section~5.3) we find that it contains long-range
interactions. Thus the condensation transition can be rationalised
within the equilibrium framework.

Moreover in the zero-range process the existence of a drive or
preferred direction, producing a conserved particle current, is not
essential for the occurrence of a condensation transition. What does
appear necessary, however, is the conservation of particles.  The
fixed number of particles implies implies the introduction of a
fugacity $z$ through (\ref{Zint}), which in turn controls the
condensation transition. As we saw the fugacity gives the hopping rate
out of a site (referred to as speed in section 5) which is a conserved
quantity. 

On the other hand, for other models the presence of a preferred
direction and conserved current does seem crucial for the existence of
phase transitions.  For example, the asymmetric exclusion process
defined in Section~\ref{Sec:nonequil} has non-trivial phase behaviour
but the undriven version (symmetric exclusion) does not.

In summary, although a general theoretical framework
for the description of phase transitions in one dimensional systems
is not yet available, we hope that the issues and models discussed
in the present paper serve to show that our understanding is developing.
\\

\noindent {\bf Acknowledgements}

Some of the work described here, particularly in Section~2.2, has been
the outcome of many enjoyable collaborations over the years and I thank
all of my collaborators warmly. Special thanks are due to Owen O'Loan
who kindly allowed me to borrow from parts of his PhD thesis in
sections 5.2 and 5.3 and with whom I have enjoyed many enlightening
discussions. I also thank Isao Hiyane for pointing  out the KDP model 
to me and Richard Blythe, Deepak Dhar and
Joachim Krug for helpful comments on the manuscript.




\end{document}